\documentclass{PoS}
\usepackage{natbib}
\usepackage{subfigure}
\usepackage{graphicx}

\title{Galaxy cluster magnetic fields from radio polarized emission}

\ShortTitle{Galaxy cluster magnetic fields }

\author{\speaker{Annalisa Bonafede}%
         \thanks{Invited talk on cluster magnetic fields.}\\
        Universit\'a di Bologna, Dip. di Astronomia, via Ranzani 1, I-40126 Bologna, Italy\\
INAF, Istituto di Radioastronomia, via Gobetti 101, I-40129 Bologna, Italy.\\
        E-mail: \email{bonafede@ira.inaf.it\\}}

\author{L. Feretti\\

INAF, Istituto di Radioastronomia, via Gobetti 101, I-40129 Bologna, Italy.}

\author{M. Murgia\\
        INAF, Osservatorio Astronomico di Cagliari, Strada 54, Loc. Poggio dei Pini, I-09012 Capoterra (Ca), Italy.}

\author{F. Govoni\\
        INAF, Osservatorio Astronomico di Cagliari, Strada 54, Loc. Poggio dei Pini, I-09012 Capoterra (Ca), Italy.}

\author{G. Giovannini\\
        Universit\'a di Bologna, Dip. di Astronomia, via Ranzani 1, I-40126 Bologna, Italy.}
\author{V. Vacca\\
         INAF, Osservatorio Astronomico di Cagliari, Strada 54, Loc. Poggio dei Pini, I-09012 Capoterra (Ca), Italy\\
Dipartimento di Fisica, Universit\'a degli studi di Cagliari, Cittadella Universitaria, 09042 Monserrato (CA), Italy.}
\abstract{The presence of magnetic fields in the intra-cluster medium
  of galaxy clusters is now well estabilished. It is directly revealed
  by the presence of cluster-wide radio sources: radio halos and radio
  relics. In the last years increasing attention has been devoted to
  the intra cluster magnetic field through the study of polarized
  radio emission of radio galaxies, radio halos and radio relics.
  Recent radio observations have revealed important features of the
  intra-cluster magnetic field, allowing us to constrain its main
  properties and to understand the physical processes taking place in
  the intra-cluster medium. I will review the newest results on galaxy
  cluster magnetic fields, both focusing on single objects and aimed
  at describing the magnetic field general properties.
 The up-coming
  generation of radio telescopes, EVLA and LOFAR, will shed light on  
several problematics regarding the cluster magnetic fields and the
  related non-thermal emission.

 . .........................  ...........................}

\FullConference{ISKAF2010 Science Meeting \\
                 June 10 -14 2010\\
                 Assen, the Netherlands}

\begin{document}

\section{Introduction}
In recent years the presence of magnetic fields in galaxy clusters has
been unambiguously proved. Magnetic fields are thought to play
an important role in the development of large-scale structure in the
Universe, but despite their importance, the origin and evolution of
magnetic fields are still open problems in fundamental physics and
astrophysics. The magnetic fields that we observe in the Local Universe
probably owe their strength to dynamo amplification of an initial
seed. The smaller the object is the shorter is the time required for
the dynamo to amplify the original seed. The magnetic fields on large
scales are thus the most challenging. The dynamical scale for large
objects are in fact long, and the amplification is correspondingly
slow. This indicates the need for a seed field whose strength is not
predicted by present theories (see {\it e.g.}
\citealt{2006AN....327..395R}).\\ According to the standard scenario
of structure formation, galaxy clusters are built-up by gravitational
merger of smaller units, such as groups and sub-clusters.  They are
composed of hundreds galaxies in Mpc-size region, and the Intra
Cluster Medium (ICM) is filled with hot and rarefied gas - emitting in
the soft-X ray domain through optically thin bremsstrahlung - magnetic
fields and relativistic particles. Magnetic fields in the ICM are
investigated through synchrotron emission of cluster-wide radio sources
and from the study of the Rotation Measure of radio galaxies. I will
discuss in this proceeding the most recent results that we have
obtained so far on the ICM magnetic field strength and structure.

\section{Why should we care about magnetic field in the ICM?}
Several models have been proposed to explain the origin of magnetic
field in the Universe.  It could be either primordial (see
\citealt{2001PhR...348..163G} for a review) or injected in the
proto-cluster region by AGN and galactic winds ejecta ({\it e.g.}
\citealt{2000ApJ...541...88V} ). Whatever the origin of the initial
seed, some amplification mechanisms are required to account for its
strength in the local Universe. Cosmological MHD simulations predict
magnetic field strength of the order of $\mu$G spread over the cluster
volume ({\it e.g.} \citealt{2005xrrc.procE8.10D}), in agreement with radio
observations ({\it e.g.} \citealt{2004IJMPD..13.1549G}). Cosmological
simulations indicate that the amplification of the magnetic field
resulting by pure adiabatic contraction of the gas falling into the
cluster potential well, can explain only 1\% of such magnetic field
strength, so that other amplification mechanisms are required.
Merger events and accretion of material onto galaxy clusters are
supposed to drive significant shear-flows and turbulence within the
ICM.  These processes can amplify the magnetic field up to at least
$\mu$G values (see \citealt{2008SSRv..134..311D} and references
therein).  A detailed knowledge of the fundamental ICM magnetic field
properties, like {\it e.g.} its power spectrum and radial decline, can help
to understand if it is amplified by processes related to merger
events.  \ \ The presence of magnetic field in the ICM of galaxy
clusters may also affect the thermal conduction: typical values for
the thermal electron gyro-radius ($\sim 10^8$ cm for T$=10^{8\circ}$K
and B$=1\mu$G) are much smaller than any scale of interest in
clusters, and in particular than the particle mean free path due to
collisions ($\sim$20 kpc). It follows that the effective mean free
path for diffusion perpendicular to the magnetic field lines is
reduced, and being the magnetic field tangled in the ICM, it is
crucial to have information at least on the magnetic field coherence
length to understand how the magnetic field inhibits the thermal
conduction.\\ In addition, a precise knowledge of the ICM magnetic
fields is also important to clarify the origin of the relativistic
particles which are responsible for the synchrotron diffuse radio
halos and relics detected in an increasing number of galaxy clusters
(see Secs. \ref{sec:halos} and \ref{sec:relics}).

\section{Power spectrum and radial profile of the ICM magnetic field}
There are several indications that the magnetic field intensity
declines with radius with a rough dependence on the thermal gas
density. This is predicted by both cosmological simulations ({\it
  e.g.}  \citealt{2005xrrc.procE8.10D}, \citealt{1999A&A...348..351D},
\citealt{2005ApJ...631L..21B}) and by the comparison between thermal
and radio brightness profiles \citep{2001A&A...369..441G}.
It is then reasonable to assume that the magnetic field declines from
the center outwards according to:
\begin{equation}
B(r)=B_0 \left[1+\frac{r^2}{r^2_c}\right]^{-\frac{3}{2} \mu}=B_0
\left[\frac{n_e}{n_0}\right]^{\eta}
\label{eq:Bprofile}
\end{equation}
with $n_e$ being the gas density distribution, and $r_c$ the
cluster core radius.\\ Faraday Rotation Measures (RM) images have often
revealed patchy structure, with RM fluctuations over a large range of
spatial scales. This indicates that the magnetic field itself
fluctuates over a range of spatial scales, and that its power spectrum
has to be considered.  
One of the simplest approaches is to consider that the magnetic field 
fluctuations are Gaussian and that their power spectrum can be 
approximated by a power-law of the form:
\begin{equation}
|B_{\Lambda}|^2\propto \Lambda^{n}
\label{eq:BPS}
\end{equation}
in a range of scales going from $\Lambda_{min}$ to $\Lambda_{max}$.
\section{Magnetic fields and radio halos}
\label{sec:halos}
The most spectacular and direct evidence that the ICM is magnetized
comes from observations of radio halos. They are wide synchrotron
emitting sources, whose emission arises directly from the ICM. Radio
halos permeate the central region of galaxy clusters, and have typical
sizes of $\sim$ 1 Mpc at 1.4 GHz. Their emission has been mainly
studied at frequencies $\nu \sim$ GHz, and only recently observations
at lower frequencies have been performed ({\it
  e.g. }\citealt{2009A&A...508...75V},\citealt{2008A&A...484..327V}).
In this frequency range, they are characterized by low surface
brightness ($\sim$ 1 $\mu$Jy/arcsec$^2$ at 1.4 GHz) and steep radio
spectrum\footnote{We define the spectrum as
  $S(\nu)\propto\nu^{-\alpha}$, with $S(\nu)$ being the flux density
  at the frequency $\nu$}, with $\alpha \geq 1$ between 300 MHz and 1
GHz \citep{2009A&A...507.1257G}. The presence of radio emission on
such large scales poses some questions about the origin of the
emitting electrons. Because of synchrotron and Inverse-Compton losses,
particles that emit around $\sim 1$ GHz in a $\mu$G magnetic field
have a life-time of $\approx 10^{8}$ yr. During this timescale they
can only diffuse for a few tens of kpc, which is very small compared
with the observed $\sim$ Mpc scale common for Radio Halos.  This
indicates that they need to be continuously injected or accelerated
across the entire cluster volume. Two main classes of models have been
proposed in the literature:\\ $\bullet$ {\it primary or
  re-acceleration models}: in which electrons are re-accelerated {\it
  in situ} through second-order Fermi mechanism by ICM turbulence
developing during cluster mergers ({\it
  e.g.}  \citealt{1977ApJ...212....1J}; \citealt{2001MNRAS.320..365B};
\citealt{2001ApJ...557..560P});\\ $\bullet$ {\it secondary or hadronic
  models}: in which electrons originate from hadronic collisions
between the long-living relativistic protons in the ICM and thermal
ions ({\it e.g.} \citealt{1980ApJ...239L..93D}).\\ We refer to
Brunetti (this conference) for a more complete analysis.  However,
because of the nature of the synchrotron radiation, it is not easy to
measure the energetic of the relativistic electrons and that of the
magnetic field using solely observations of radio halos. The minimum
energy calculation is problematic and the detections of non-thermal
inverse Compton radiation in the Hard X-ray band are difficult and
limited to few clusters. Indeed, an independent method to estimate the
magnetic field properties in the ICM is required (see Sect.6).\\ Radio
halos are usually found to be un-polarized. This could be due to the
low resolution that is often needed to properly image their emission,
causing beam depolarization, and/or to internal depolarization caused
by differential Faraday Rotation (see Sec. \ref{sec:haloDep}). In the
Coma cluster upper limits to the fractional polarization are
$\sim$10\% at 1.4 GHz, and lower values ($\sim$6\% and 4\%) have been
found for two other powerful halos in Abell 2219 and Abell 2163 (see
\citealt{2004IJMPD..13.1549G} and references therein). Polarized
emission from radio halos has been observed so far in the cluster
Abell 2255 (\citealt{2005A&A...430L...5G}, see also \citealt{PizzoTesi}) and MACS\,J0717+3745 \citep{2009A&A...503..707B}.
\begin{figure*}[t] 
\centering
{\includegraphics[width=\textwidth]{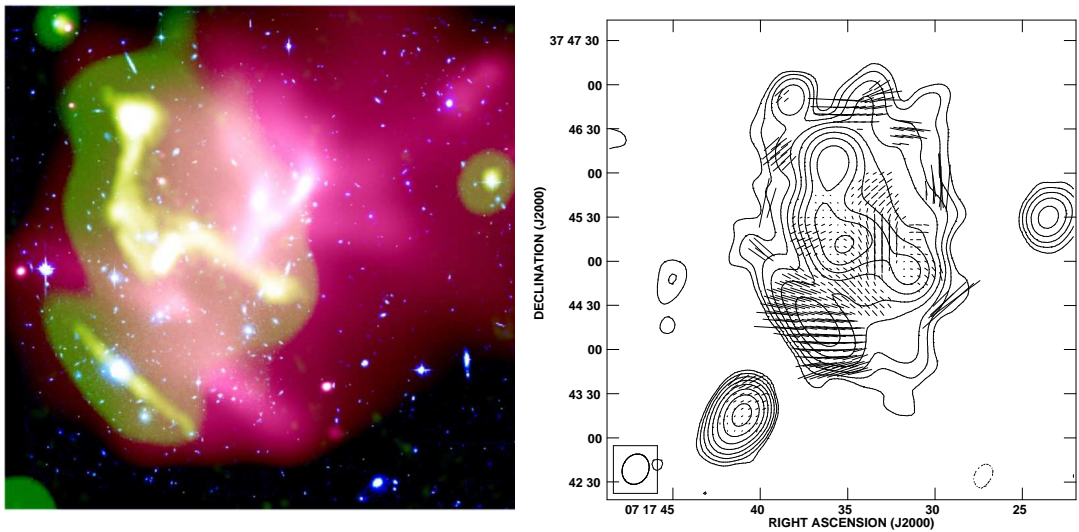}}
\label{fig:MACS}
\caption{The most powerful radio halo known so far in MACS\,J0717+3745
  Left panel: Optical emission from HST in blue, X-ray emission from
  Chandra (0.5-7 keV band) in magenta and radio emission (VLA at 1.4
  GHz) in green. X-ray and optical images are courtesy of
  C.J. Ma. Right panel: Contours show to the total-intensity emission
  (beam is 21$''\times$18$''$) starting at 3$\sigma$, and spaced by
  factors of 2. Lines refer to the E vectors. Their orientation
  represents the projected E-field. Their length is proportional to
  the fractional polarization: 1$''$ corresponds to 13\% (Fig. from
  \citet{2009A&A...503..707B}) .}
\label{fig:MACS}
\end{figure*}
\subsection{Polarized emission from radio halos}
\label{sec:halosPol}
Detecting polarized emission from cluster radio halos would be
extremely important for the study of the magnetic field power spectrum
and morphology. \citet{2004A&A...424..429M} developed the FARADAY code
to simulate 3D magnetic fields in galaxy clusters. They investigated
how different magnetic field power spectra affect the shape and the
polarization properties of radio halos (see Fig. 6 of that
paper). When single power-law power spectra are considered, models
with $n>$3 and $\Lambda_{max}>$100 kpc (see Eq. \ref{eq:BPS}) result
in magnetic fields whose energy is larger on the large spatial scales,
thus giving rise to filamentary and polarized radio halos. Models with
$n<2$, instead, having most of the magnetic field energy on small
spatial scales, will give rise to halos with a more regular
morphology, and very little polarization.\\ By comparing simulations
and observations, the detection of polarized emission from radio halos
permits to put some constraints on the magnetic field power
spectrum. \citet{2006A&A...460..425G} analyzed the polarized emission
from the radio halo in the cluster Abell 2255. They found that a
single power law cannot account for the observed polarization. A power
spectrum with spectral index $n=2$ at the cluster center and $n=4$ at
the cluster periphery is needed to produce the observed polarized
emission. Comparing observations and numerical simulations
\citet{2009A&A...503..707B} found that the magnetic field power
spectrum in the cluster MACS\,J0717+3745 should be steeper than 3, and
that its maximum scale $\Lambda_{max}$ should be $>$100 kpc. It is
interesting to note that the Kolmogorov power spectrum, in this 3-D
notation, would have a slope $n=11/3$. We note that the theory
developed by Kolmogorov treats incompressible un-magnetized and
uniform media, so that its application to the case of the ICM of
galaxy clusters is all but obvious.  Nonetheless, observational data
and cosmological simulations indicate that it is a good description of
the pseudo-pressure fluctuations \citep{2004A&A...426..387S}, velocity
field \citep{2009A&A...504...33V} and magnetic field
(\citealt{2009arXiv0912.3930K}, \citealt{2008A&A...483..699G},
\citealt{2005A&A...434...67V}, \citealt{2010A&A...513A..30B}) in the
ICM.  In Fig. \ref{fig:MACS} the cluster MACS\,J0717+3745 and the
polarized emission from the radio halo are shown.
\subsection{Un-polarized radio halos: internal and external depolarization}
\label{sec:haloDep}
There are different possibilities to explain why polarized emission
from radio halos is so rarely detected. Radio halo emission fills the
central Mpc$^3$ of galaxy clusters. Here magnetic fields, relativistic
electrons and thermal gas coexist and interact. It is reasonable to
assume that the direction of the magnetic field varies from point to
point in the ICM. The intrinsically polarized radiation from
relativistic electrons, will then vary accordingly. In addition, when
radio waves propagate through the ICM, the polarization plane is
subject to the internal Faraday Rotation. The result of these effects
is the well known frequency dependent depolarization of the intrinsic
synchrotron emission. Furthermore, radio halos are faint and wide
sources, so that low resolution observations are often required to
properly image their emission. If differential Faraday rotation is
occurring within the beam, {\it i.e.} if the minimum scale of the
magnetic field power spectrum is smaller than the beam FWHM, the
so-called ``beam-depolarization'' is expected to be particularly
strong. Recent works on RM studies indicate that magnetic field may
fluctuate on scales as small as few kpc, while typically radio halos
are observed with a resolution of several tens to hundred
kpc.\\ Recently, \citet{2010A&A...514A..71V} have analyzed the radio
halo in the cluster Abell 665, where polarized emission is not
detected.  By comparing observed and synthetic radio halo images the
strength and structure of the intra-cluster magnetic field has been
constrained. The simulated magnetic field that best reproduces the
observations in Abell 665 is characterized by $B_0=$1.3 $\mu$G,
$\eta=$0.47. Once a Kolmogorov power spectrum is assumed, spatial scales
$\Lambda$ are up to 450 kpc. The authors investigated the
different causes (noise and resolution) that produce depolarization
(see Fig. 3 of the paper). The simulated radio halo appears intrinsically  polarized
at 1.4 GHz, with fractional polarization of $\sim$24 \%, indicating
that internal depolarization alone cannot explain why the radio halo
appear to be un-polarized. When the intrinsic radio emission is
observed with a resolution of $\sim$ 15$''$, the polarization is
reduced at $\sim$7\%. But when noise, typical of observations
performed with present instruments ($\sim$ 15 $\mu$Jy/beam) is added,
the polarized emission falls below the detection level. This
illustrates that the intrinsic polarization of radio halos is not
completely canceled out neither by internal depolarization nor by beam
effects. Radio halos would appear polarized if observed with enough
sensitivity, but detecting the polarized emission is a very hard task
with current interferometers. The radio halo in MACS\,J0717 +3745 is
indeed the most powerful radio halo observed so far. The mean
fractional polarization is $\sim$5 \% at 20 cm
\citep{2009A&A...503..707B} that is not much below the upper limits
found for other radio halos.\\ The new generation of radio telescopes,
such as LOFAR and EVLA, will have the chance to detect polarized
emission from radio halos thanks to their better sensitivity
performances. As shown by Govoni (2007, contributed talk at LOFAR
meeting in Emmen, NL), LOFAR will have the chance of detecting
polarized emission in the higher frequency band ($\sim$ 200
MHz).
\begin{figure*}[htb] 
\centering
\includegraphics[width=\textwidth]{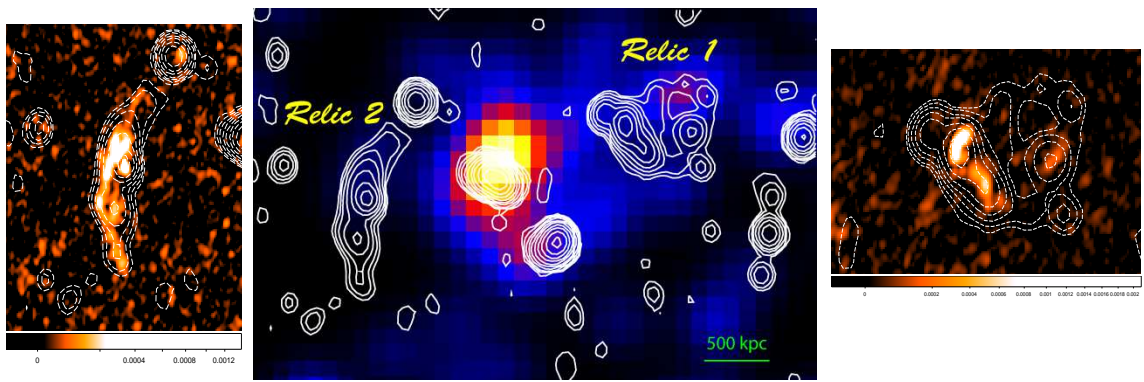}
\caption{Center: Abell 2345 X-ray emission (colors) in the energy band
  0.1-2.4 keV from RASS; contours represent the radio image of the
  cluster at 1.4 GHz. The beam is 50$''$$\times$38$''$. Contours are
  0.24 mJy/beam and are then spaced by a factor 2. Side panels: colors
  refer to the polarized emission at 1.4 GHz (beam FWHM $23''\times
  16''$), contours are like in the central panel.}
\label{fig:A2345}
\end{figure*}
\section{Polarized emission from radio relics}
\label{sec:relics}
Another indication that the ICM is magnetized comes from the
observations of the so-called ``radio relics''. Their size ($\sim$
Mpc), surface brightness, and radio spectrum have properties similar
to radio halos, but they differ in morphology and polarization
properties. They are located at the outskirts of the host galaxy
clusters, usually at the boundary of the X-ray emission, and are
strongly polarized, with linear fractional polarization at 1.4 GHz of
10-30 $\%$, reaching values up to 50 $\%$ in some regions (see {\it
  e.g.} \citealt{2009A&A...494..429B}). Their morphology is quite
various, but it is usually elongated.\\ The origin of their emission
is still uncertain, and different models have been proposed in the
literature:\\
$\bullet$ ``Radio-ghost''\footnote{We use the
  nomenclature proposed by \citet{2004rcfg.proc..335K}}: aged radio
plasma revived by merger or shock wave through adiabatic compression
(see \citealt{2001A&A...366...26E});\\ $\bullet$ ``radio gischt'':
electrons are accelerated by Fermi-I process (Diffusive Shock
acceleration) caused by merging or accretion shock, and the magnetic
field is amplified (\citealt{1998A&A...332..395E},
\citealt{2007MNRAS.375...77H}).\\ The ``Radio-ghost'' model predicts a
curved radio spectrum, a filamentary or toroidal morphology and
magnetic field aligned with the filamentary structure, while ``radio
gischt'' models predict a straight radio spectrum, and an arc-like
morphology of the relic, with magnetic field aligned to the relic main
axis. The presence of double relics in a single cluster is
particularly interesting in the ``radio gischt'' scenario, since the
shape, morphology and properties of these extended structures strongly
suggest the presence of shock waves propagating from the cluster
center to the peripheral regions. A few clusters with double relics are
known so far, and a detailed study of their radio properties,
including all the predictions done by the models, has not been
systematically performed yet. \\ Recently, we have analyzed the double
relics in clusters Abell 1240 and Abell 2345
\citep{2009A&A...494..429B}, and studied their spectral index,
spectral index trend and polarization properties.  Double relics in
Abell 1240 show radio morphology, spectral index and polarization
values in agreement with radio gischt predictions in the case of
merger shock waves. The values of the Mach numbers indeed ($\sim$2-3)
disfavour the hypothesis of accretion shock waves. The same properties
characterize also one of the two relics in Abell 2345 (relic 2 in
Fig. \ref{fig:A2345}). Here the polarized emission is quite
spectacular, tracing the arc-like structure morphology of the
relic. Relic 1 in A2345 has instead peculiar features. It shows a
peculiar morphology and un-expected spectral index profile, with
steeper values toward the cluster outskirts. The X-ray emission of
Abell 2345 shows multiple substructures that could be galaxy groups
interacting with A2345. Peculiar features of Relic 1 could be
explained by a shock wave moving inward, due to the interaction of the
main cluster with another group. Recently optical analysis of this
region \citep{2010arXiv1007.0848B} have been performed, confirming
this possible scenario. Deep X-ray observations would be required to
better understand this issue.
\subsection{Why are relics so rare?}
Despite all models agree in saying that shock waves are responsible
for relic formation, yet temperature and brightness gradients induced
by shock waves have been unambiguously found only in A520
\citep{2005ApJ...627..733M}, and in the Bullet cluster
\citep{2002ApJ...567L..27M}, while in the case of the Coma cluster,
\citet{2006A&A...450L..21F} did not find any evidence of a temperature
jump near the Coma cluster relic. A few more cases are good
candidates, but due to low X-ray brightness at the cluster periphery
searching for shock associated to radio relics is a very hard task.
Recently, \citet{2010arXiv1003.5658V} have performed high resolution
($\sim$ 25 kpc/h) cosmological simulations of a sample of 20 massive
galaxy clusters, at the aim of studying shocks statistics in the
cluster formation region (Vazza et al., in prep). They found that the
occurrence of shocks with $M>2$ within the virial radius ($R_v$) is a rare
event, and only two clusters over 20 are interested by strong
shocks inside $R_v /2$ at z $=$ 0, in agreement with X-ray
observations.
\section{Faraday Rotation measures and ICM magnetic fields}
\label{sec:RM}
In the last years new important results on the ICM magnetic fields
have been obtained by analysing the Faraday RM of radio
galaxies in/behind clusters. Synchrotron radiation from radio
galaxies that crosses a magneto-ionic medium, as the ICM, is subject
to the Faraday Rotation. The direction of the polarization plane:
$\Psi_{int}$ is rotated of a quantity that in the case of a purely
external Faraday screen is proportional to the square of the
wavelength:
\begin{equation}
\Psi_{obs}(\lambda)=\Psi_{int}+RM\lambda^2; RM=\int_0^L{B_{//}n_edl};
\label{eq:psiobs}
\end{equation}
here $B_{//}$ is the magnetic field component along the line of sight,
$n_e$ is the ICM gas density and $L$ is the integration path along the
line of sight. With the help of X-ray observations, providing
information on $n_e$ distribution, RM studies give an additional set
of information about the magnetic field in the ICM.
\subsection{The Coma cluster magnetic field} 
\begin{figure*}[htb] 
\centering
\subfigure{\includegraphics[width=0.45\textwidth]{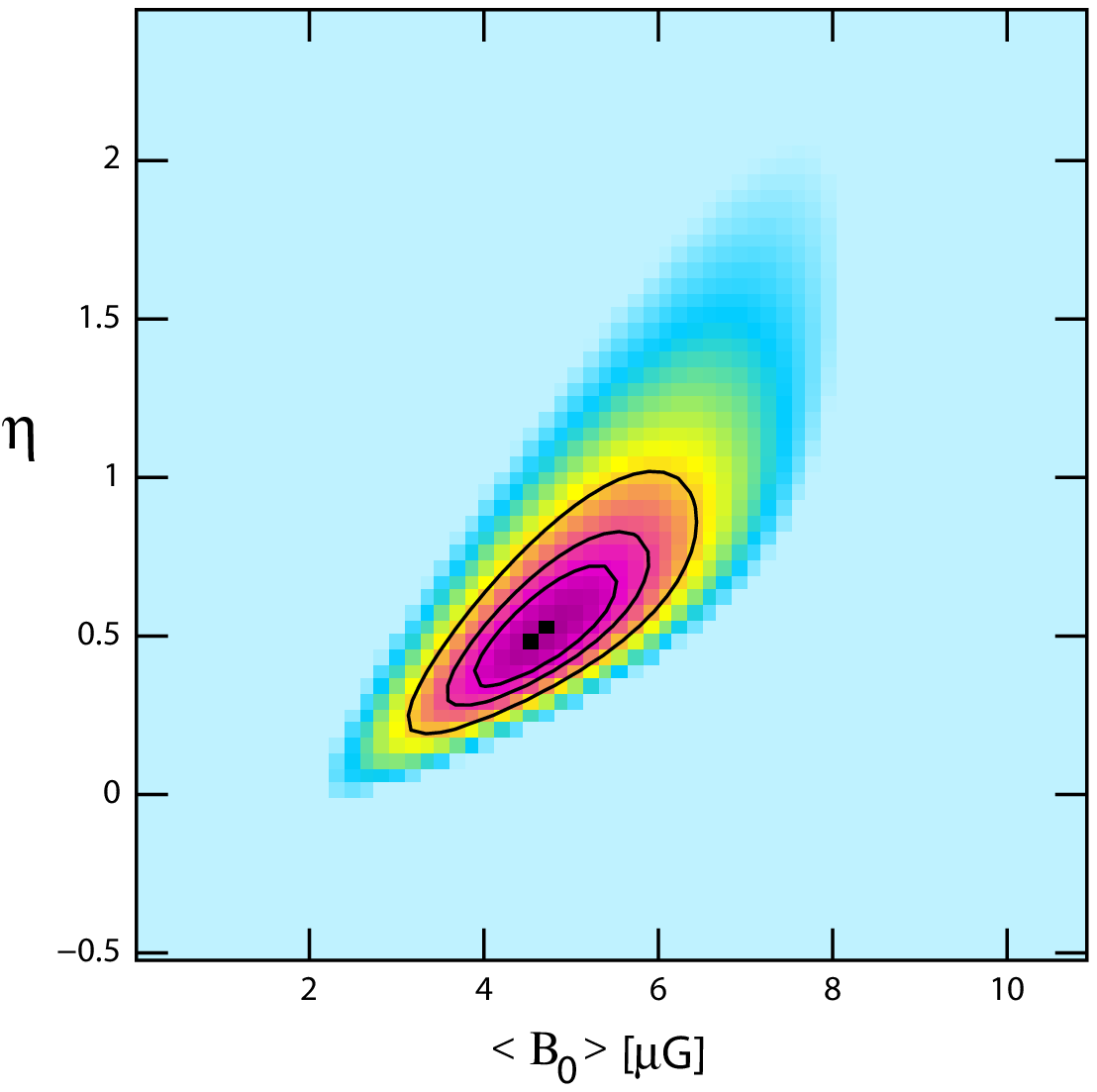}}
\subfigure{\includegraphics[width=0.45\textwidth]{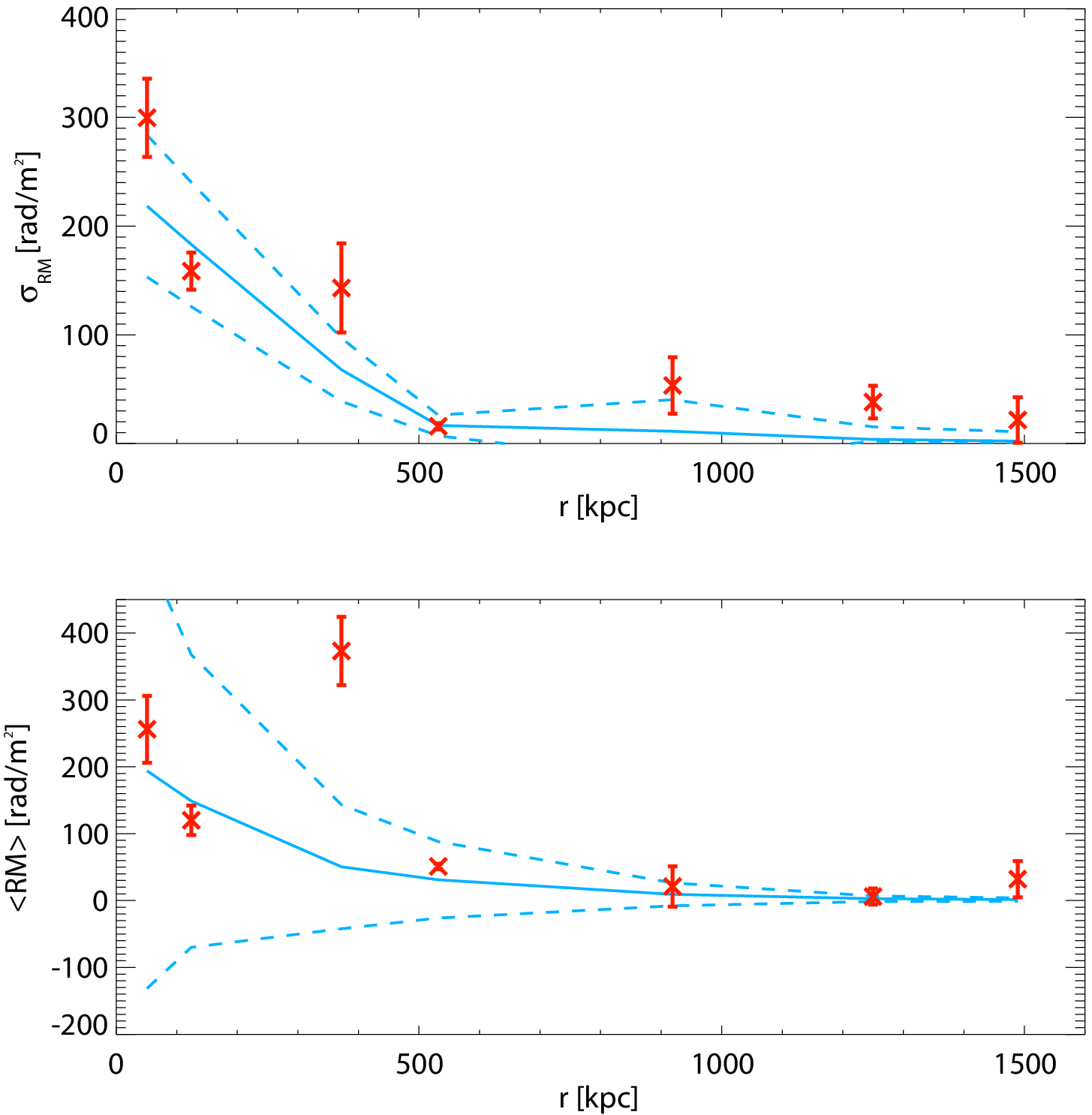}}
\caption{Figure from \citet{2010A&A...513A..30B}. Left panel: $\chi^2$
  plane obtained by comparing simulated and observed RM
  images. Contours refer to 1,2 and 3-$\sigma$ confidence
  levels. Right panel: RM mean and dispersion for the best-fit
  magnetic field model (cyan), observed points are shown in red. }
\label{fig:COMA}
\end{figure*}

We have recently analysed the RM images of 7 extended sources in the
Coma cluster. Observations were performed at 4 or 5 frequencies in the
range 1.4 - 8.9 GHz, with angular resolution of $\sim$ 1.5$''$,
corresponding to 0.7 kpc at the Coma redshift. The RM images we
have obtained allowed us to constrain the magnetic field radial
decline with unprecedent detail.\\ We used the approach proposed by
\citet{2004A&A...424..429M} and simulated RM images for different
magnetic field models with different values of $B_0$, going from 1 to
11 $\mu$G, and $\eta$ going from -0.5 to 2.5. In Fig. \ref{fig:COMA}
the $\chi^2$ plane obtained for each combination of these parameters
is shown. Due to the degeneracy between the two parameters, different
models give similar $\chi^2$. The magnetic field radial profile is
better described by a model with $B_0=$4.7$\mu$G and $\eta=$0.5,
indicating that the magnetic field energy density follows the gas
energy density. It is interesting to note that also the model with
$\eta=0.67$, expected in the case of a magnetic field frozen into the
gas, lies within the 1-$\sigma$ confidence level of the $\chi^2$.
Magnetic field models with a profile flatter than $\eta < 0.2$ and
steeper than $\eta > 1.0$ are instead excluded at 99\% confidence level, for
any value of $\langle B_0 \rangle$. \\ 
The value derived here can be compared with other estimates derived from
equipartition and Inverse Compton hard X-ray emission. The magnetic
field model resulting from our RM analysis gives an average magnetic
field strength of $\sim$ 2 $\mu$G, when averaged over the central
$Mpc^3$. Equipartition estimate is strongly dependent on the unknown
particle energy spectrum, and a single power-law is often
assumed. Although the many uncertainties and assumptions relying under
this estimate, the model derived from RM analysis is compatible with
the equipartition magnetic field (0.7 - 1.9 $\mu$G;
\citealt{2003A&A...397...53T}). A direct comparison with the magnetic
field estimate derived from the IC emission is more difficult, since
the Hard-X detection is debated, and depending on the particle energy
spectrum, the region over which the IC emission arises may change. The
model derived from RM analysis gives a magnetic field estimate that is
consistent with the present lower limits obtained from Hard X-ray
observations by {\it e.g.} \citet{2009ApJ...696.1700W}, while it is a
factor four higher than the value derived from
\citet{2004Ap&SS.294...37F}.\\ The knowledge of the ICM magnetic field
breaks the synchrotron degeneracy and permits to constrain the energy
content of electrons that give rise to radio halo emission. Using the
result of RM studies, \citet{2010MNRAS.401...47D} have investigated
through cosmological simulations if hadronic models may account for
the observed radio emission in the Coma halo. The authors found that
the radial profile of the radio halo can only be reproduced with a
radially increasing energy fraction within the cosmic ray proton
population, reaching $>$100\% per cent of the thermal-energy content
at 1 Mpc from the cluster center, thus disfavoring the hadronic
models.
\subsection{Magnetic field power spectrum from Faraday RM}
The intra cluster magnetic field power spectrum can be constrained
starting from RM images of extended radio source, both within
and behind galaxy clusters. In the cluster Abell 2255
\citep{2006A&A...460..425G} the power spectrum has been analyzed also
thanks to the detection of polarized filaments, as discussed in
Sec. \ref{sec:halos}. In the Coma cluster \citep{2010A&A...513A..30B}
the power spectrum was assumed to be Kolmogorov-like, and by comparing
the auto-correlation function and the structure function of simulated
and observed RM images, the maximum and minimum scale of the power
spectrum were constrained to be $\sim$ 2 and 34 kpc
respectively. Similarly, the power spectrum of Abell 2382 is found to
be consistent with a Kolmogorov-like power spectrum, with scales going
from few to 30 kpc. \\ A different approach was adopted by
\citet{2009arXiv0912.3930K} and \citet{2005A&A...434...67V} that
derived the magnetic field power spectrum through Bayesian analysis of
RM images in the clusters HydraA, Abell 400 and Abell 2634. In these
clusters the power spectrum is found to consistent with a Kolmogorov
slope over a wide range of spatial scales.
\begin{figure*}[htb] 
\centering
\subfigure{\includegraphics[width=0.45\textwidth]{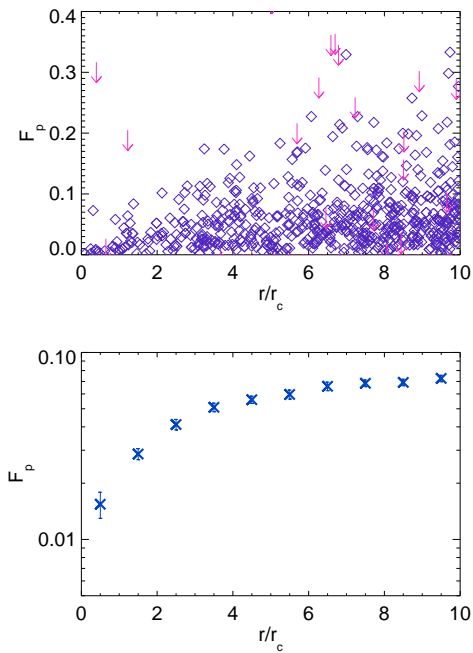}}
%\subfigure{\includegraphics[width=0.42\textwidth]{comboPOL.ps}}
\caption{Fractional polarization for the whole cluster sample. Upper
  panel: values from single sources; lower panel: average values in
  bins having width of one core radius. Preliminary results from
  Bonafede et al. (in prep). }
%Right panel: Fractional
%  depolarization trend for the radio quiet (top) and radio halo
%  (bottom) sub-samples.}
\label{fig:depo}
\end{figure*}
\section{Depolarization from radio sources}
\label{sec:depo}
The presence of ICM magnetic fields affects the polarization
properties of radio galaxies. The higher RM suffered by sources in the
inner regions, where both magnetic field strength and gas density are
higher, causes in fact high beam depolarization. Sources at larger
radii, instead, are subject to less depolarization, since their
emission is affected by lower RM.\\ At the aim of investigating the general
properties of the ICM magnetic field, we selected massive galaxy
clusters from the HIghest X-ray FLUx Galaxy Cluster Sample
\citep{2002ApJ...567..716R}, and used Northern VLA Sky Survey data to
analyze the polarization properties of radio-sources out to 10 $r_c$
from the cluster centers (Bonafede et al., in prep). The sample
consists of 33 clusters, with $L_x[0.1-2.4 keV]\geq 1.5 \times
10^{44}$erg/s. In Fig. \ref{fig:depo} the
fractional polarization as a function of the distance from the
cluster center is shown. We detected a clear trend of the
fractional polarization, being smaller for sources close to the
cluster center and increasing with increasing distance form the cluster
central regions. This confirms, as already found by
\citet{2004JKAS...37..337C} and \citet{2004mim..proc...13J} that
magnetic fields are ubiquitous in galaxy clusters.\\ The sample of
clusters we have selected comprises both clusters that host radio
halos and clusters that are radio quiet. We searched for possible
differences in the magnetic field properties in clusters with and
without radio halos by comparing the depolarization trend for the two
sub samples. The Kolmogorov-Smirnov
statistical test gives high probability  that the
polarization trend observed in the radio-halo and radio-quiet sample are
taken from the identical intrinsic population. Magnetic fields in
galaxy clusters are then likely to share the same properties
regardless of the presence of radio emission from the ICM. This result
poses problems to the ``hadronic-models'' for the origin of radio
halos, that requires a net difference in clusters with and without
radio halos, while is in agreement with the re-acceleration scenario (see Brunetti, this conference).
\section{Conclusions}
Magnetic fields are now recognized to be an important component of the
ICM in galaxy clusters. Our knowledge of their properties
has greatly improved in the recent years thanks to both radio
observations and the developments of new techniques to interpret
data. We have shown that deep radio observations of both radio
galaxies and cluster-wide diffuse sources permit to constrain the
magnetic field radial decline slope and power spectrum.  The results
obtained by radio studies are fundamental to understand the process of
radio halo and radio relic formation. The new generation of radio
interferometers, such as LOFAR, that will open an unexplored frequency
window, and as the EVLA, that will largely improve the sensitivity and
polarization performances at $\nu>1$ GHz, will allow us in the next
years a step forward in the knowledge of the ICM magnetic field.\\

\small{
{\it Acknowledgements} A.B. thanks the conference SOC for the
invitation to this very interesting conference. We are also
grateful to our collaborators: K. Dolag and G. Taylor for useful
discussions.}

%\begin{thebibliography}{99}
%  \bibitem{...} ....
%\end{thebibliography}
\small{
\bibliographystyle{aa} \bibliography{/Locale/DEPO/PAPER/master}}

\end{document}